\documentclass{article}
\usepackage{amsmath}
\usepackage{amssymb}
\usepackage{cite}

\begin{document}

\newcommand{\pst}{\hspace*{1.5em}}

\newcommand{\be}{\begin{equation}}
\newcommand{\ee}{\end{equation}}
\newcommand{\bm}{\boldmath}
\newcommand{\ds}{\displaystyle}
\newcommand{\bea}{\begin{eqnarray}}
\newcommand{\eea}{\end{eqnarray}}
\newcommand{\ba}{\begin{array}}
\newcommand{\ea}{\end{array}}
\newcommand{\arcsinh}{\mathop{\rm arcsinh}\nolimits}
\newcommand{\arctanh}{\mathop{\rm arctanh}\nolimits}
\newcommand{\bc}{\begin{center}}
\newcommand{\ec}{\end{center}}
\def\vep{\varepsilon}

\begin{center} {\Large \bf
\begin{tabular}{c}
COMPARING ENERGY DIFFERENCE AND FIDELITY 
\\[-1mm]
OF QUANTUM STATES
\end{tabular}
 } \end{center}


\begin{center} {\bf
Victor V. Dodonov
}\end{center}


\begin{center}
{\it
 Instituto de F\'{\i}sica, Universidade de Bras\'{\i}lia, 
Caixa Postal 04455, 70910-900 Bras\'{\i}lia, DF, Brazil
}

 e-mail:~~~vdodonov@fis.unb.br\\
\end{center}

\begin{abstract}\noindent
We look for upper bounds of the relative energy difference of two pure quantum states  
with a fixed fidelity between them or upper bounds of the fidelity for a fixed relative energy difference.
The results depend on the concrete families of states chosen for the comparison.
Exact analytical expressions are found for several popular sets of states: coherent, squeezed vacuum,
binomial, negative binomial, and coherent phase states. Their consequence is that
 to guarantee, for example, the relative energy difference less than 10\% for quite arbitrary (unknown)
coherent states, the fidelity  must exceed the level $0.995$. For other kinds of states, the
restrictions can be much stronger.  \end{abstract}


\noindent{\bf Keywords:}
fidelity, energy difference, harmonic oscillator, superpositions of Fock states, coherent states, squeezed states,
binomial and negative binomial states, phase coherent states, ``hyper-Poissonian'' states.

\section{Introduction}
\pst
The quantification of closeness between different quantum states 
is important for many applications of quantum mechanics to the quantum information theory, such as
quantum tomography, quantum teleportation, quantum states engineering, and so on.
A very popular quantity, which can be encountered in almost every contemporary paper on quantum information,
is the so called {\em fidelity\/} introduced in \cite{Joz}. 
 For {\em pure\/} states, which are the only subject of this paper,
this quantity is nothing but  the square of the absolute value of the scalar product
 between two states $|\psi_1\rangle$ and $|\psi_2\rangle$:
\be
{\cal F}=\left|\langle\psi_1|\psi_2\rangle\right|^2.
\label{F}
\ee
Of course, this quantity is well known and was used long before the fancy name was given for it
\cite{Barg,Woot,Braun}.

The question I am interested in is as follows: how different can two quantum states 
with a given value of the fidelity between them be?
For example, one can find in the literature various boundaries or critical values for fidelities, corresponding to different 
teleportation protocols of some specific classes of quantum states. In particular, the value ${\cal F}=1/2$
has been established as the boundary between classical and quantum domains in the teleportation of
coherent states of the electromagnetic field \cite{Braun00}. Another critical value ${\cal F}=2/3$ was found in
\cite{Cerf00,Gran01}. The meaning of these numbers was further elucidated in \cite{Ban04,Cav04}.
Note that the first reported experimental values were rather low: between $0.56$ and $0.66$ 
\cite{Furu98,Bow03,Zhang03}. 
Only recently the levels exceeding $90$\% were attained \cite{Zavat09,Specht11}.

But is it sufficient to have the fidelity of, say, $95$\% or even $99$\%, 
to be sure that the two states are ``really'' close to each other?
Of course, the answer depends on the concrete situation and additional information or assumptions about the states,
in particular, on the exact meaning of the word ``close''. 
If it is known that parameters
of the two states can vary in some restricted intervals only, then even not very high levels of fidelity can be sufficient sometimes.
But in the generic case this is definitely not so. 

The following example clearly shows the essence of the problem to be studied in this article.
Let us compare a superposition of two Fock states $|\psi_1\rangle =|n\rangle$ and
$|\psi_2^{(m)}\rangle = \sqrt{1-|\beta|^2}|n\rangle + \beta|m\rangle$ (with $|\beta|\le 1$).
The fidelity between the two states equals ${\cal F}= 1-|\beta|^2$, and it does not depend on the value $m$.
Suppose that, for definiteness, $n=1$ or $n=0$.
The question is: are the states $|\psi_2\rangle$ with $m=2$ and $m=100$ ``equally distant'' from $|\psi_1\rangle$
or not? It seems that the answer depends on the intuition or, more likely, on some additional assumptions
related to the concrete physical system under study. If the accepted answer is {\em yes\/}, then the subject
is closed, and there is no need to continue reading.
But if an intuition tells somebody that the state $|\psi_2^{(100)}\rangle$ is ``more remote'' from $|\psi_1\rangle$ 
than $|\psi_2^{(2)}\rangle$, then some interesting further problems can be formulated and resolved.
\footnote{In a sense, this situation resembles that with the problem of phase in quantum mechanics.
It is known that a well defined phase operator satisfying {\em all\/} reasonable requirements does not exist.
Nonetheless, there exists a tremendous literature on the subject, and many interesting results were obtained.}

First of all, we need some additional physical quantity, which could permit us to distinguish between the states
$|\psi_1\rangle$ and $|\psi_2\rangle$. Of course, its choice is by no means unique,
so I choose the simplest one (rather natural, from my point of view): the energy. Then one can introduce,
besides the usual distance in the Hilbert space, based on the scalar product, a ``polarized distance''
(or ``energy-sensitive'' distance) \cite{dod99}, which shows that $|\psi_2^{(100)}\rangle$ is more 
``far away'' from $|\psi_1\rangle$ than $|\psi_2^{(2)}\rangle$. Here we wish to find some relations between
the {\em energy difference\/} of two states and their fidelity. All the results are related to the
one-dimensional harmonic oscillator (or a single mode of the electromagnetic field), and all the quantities
are assumed dimensionless, i.e., I take formally $\hbar=\omega=m=1$, where $\omega$ and $m$ are the frequency and
mass of the oscillator.
The difference of the mean energies in the states $|\psi_1\rangle$ and $|\psi_2\rangle$ equals 
\be
\Delta E\equiv E_2 - E_1= \beta m -\left(1- \sqrt{1-|\beta|^2}\right)n.
\label{mn}
\ee
This simple formula shows two remarkable results. First, it becomes obvious that for {\em any\/}
fixed value of the fidelity ${\cal F}$ one can always find two states with an {\em arbitrarily large\/}
energy difference $\Delta E$, provided that no restrictions on the choice of states are imposed.
However, less trivial and, perhaps, more interesting results arise if one considers only some fixed
subset of all possible states. Namely such situations happen, as a matter of fact, in the analysis
of real experimental data, when researchers suppose from the very beginning that the states they study 
belong to some fixed family (e.g., coherent states, squeezed states, Gaussian states, and so on).
My goal is to consider some interesting known families of states and to find
the maximal possible value of $\Delta E$ for the fixed ${\cal F}$,
or, equivalently, maximal possible value of ${\cal F}$ for the fixed  $\Delta E$,
 {\em within the given family of states\/}. The concrete states considered in this paper
 are: coherent states, squeezed vacuum states, negative and positive binomial states, and their
special cases. 

A more restricted task can be to find mutual limitations on 
$\Delta E$ and $\vep \equiv 1-{\cal F}$ in the cases when both these quantities are {\em small\/}
(perhaps, this is the most important special case from the practical point of view). Here formula (\ref{mn})
also gives an insight --- considering the case $|\beta|\ll 1$ (i.e., $\vep \ll 1$) 
and limiting the Hilbert space (the numbers
$m$ and $n$), one obtains the relations
\be
(\Delta E)_{max} = \gamma\sqrt{1 -{\cal F}}, \qquad {\cal F}_{max} = 1- (\Delta E)^2/\gamma^2,
\label{gamma}
\ee
where $\gamma$ is some constant value (which depends on $m$, i.e., on the chosen family of states).
Note that relations (\ref{gamma}) are typical for many sets of states.
Indeed, a standard way of constructing the families of states is to take some ``fiducial'' state
$|\psi_0\rangle$ and to act upon it by unitary operators depending on some parameter, creating
the states \cite{Klaud,Perel}
$
|\psi_s\rangle =\exp\left(is\hat{\cal O}\right)|\psi_0\rangle$ 
(with $\hat{\cal O}=\hat{\cal O}^{\dagger}$).
Well known examples are coherent states, introduced in this way 60 years ago 
  \cite{Feyn51,Glaub51,Fried} (without knowing this name yet), and squeezed states,
introduced in a similar way for the first time approximately at the same period \cite{Fried,InfPleb}.
For $s\to 0$ one has (similar relations were found in different contexts, e.g., in 
\cite{Anan90,Abe93})
\be
{\cal F} = 1- s^2\left(\langle\psi_0|\hat{\cal O}^2|\psi_0\rangle -
\langle\psi_0|\hat{\cal O}|\psi_0\rangle^2 \right), \qquad
\Delta E = is \langle\psi_0|\left[\hat{H},\hat{\cal O}\right]|\psi_0\rangle,
\label{FEs}
\ee
where $\hat{H}$ is the Hamiltonian of the system. Then the constant $\gamma$ can be found by maximizing
$\Delta E$ or ${\cal F}$ with respect to all states $|\psi_0\rangle$ belonging to the chosen family.
Only in rare cases, when the expansions in (\ref{FEs}) start from the higher-order terms, the relations
(\ref{gamma}) should be replaced by other ones.
It is impressive that in all the cases studied in this paper the upper bounds for the fidelity can be
obtained in explicit analytical forms, from which the constant $\gamma$ in (\ref{FEs}) can be calculated
by simple expansions in the Taylor series.

\section{Coherent states}
\pst
Let us start with coherent states, defined as
\be
|\alpha_j\rangle =\exp\left(\alpha_j \hat{a}^{\dagger}-\alpha_j^*\hat{a}\right)|0\rangle
=\exp\left(-|\alpha_j|^2/2\right)\sum_{n=0}^{\infty}\frac{\alpha_j^n}{\sqrt{n!}}|n\rangle,
\label{coh}
\ee
where $\hat{a}$ and $\hat{a}^{\dagger}$ are standard bosonic annihilation and creation operators,
$\left[\hat{a},\hat{a}^{\dagger}\right] =1$. The fidelity and the energy difference between two states
$|\alpha_1\rangle$ and $|\alpha_2\rangle$ are well known:
\be
{\cal F} =\exp\left(-\left|\alpha_2-\alpha_1\right|^2\right), \qquad
\Delta E = E_2 -E_1 = \left|\alpha_2\right|^2-\left|\alpha_1\right|^2.
\label{cohFE}
\ee
Since the energies depend on the absolute values of $\alpha_j$ only, the maximal fidelity for the fixed
$\Delta E$ is achived if the phases of complex numbers $\alpha_2$ and $\alpha_1$ coincide. Therefore
hereafter I assume that each $\alpha_j$ is real and $\alpha_1 \ge 0$. 
It is easy to see that for the fixed value of ${\cal F}$
the energy difference can be as big as desired if the values $\alpha_j$ are big enough. 
Therefore it seems reasonable to analyze the {\em relative energy difference\/}
\be
{\cal E} \equiv \left(E_2 -E_1\right)/E_1 = \frac{\alpha_2^2 -\alpha_1^2}
{1/2 +\alpha_1^2},
\label{defcalE}
\ee
which is certainly limited (the contribution of the ground state energy $1/2$ is essential
here, because dividing $\Delta E$ simply by the mean photon number 
$\langle \hat{a}^{\dagger}\hat{a}\rangle$ one cannot eliminate very big values for the states
close to the vacuum).
Resolving equation (\ref{defcalE}) with respect to $\alpha_2$ and putting the positive root
(as the most close to $\alpha_1$) in (\ref{cohFE}), one arrives at the problem of finding
the extremal value of the function 
$f(\alpha_1) = \sqrt{\alpha_1^2(1+{\cal E}) +{\cal E}/2} -\alpha_1$ for the fixed value ${\cal E}$.
This extremum (positive maximim for ${\cal E}>0$ and negative minimum for ${\cal E}<0$)
is achived for $\alpha_1^2=1/[2(1+{\cal E}]$, and the final result for the {\em maximal possible
fidelity\/} between two coherent states with the fixed relative energy difference ${\cal E}$ is
\be
{\cal F}_{max}^{(coh)}= \exp\left(-{\cal Y}^2/2\right),
\label{Fcohmax}
\ee
where the new quantity ${\cal Y}$ is nothing but the {\em symmetrical\/} relative energy difference:
\be
{\cal Y} =\frac{|{\cal E}|}{\sqrt{1+{\cal E}}} \equiv \frac{|E_2 -E_1|}{\sqrt{E_2E_1}}.
\label{defY}
\ee
Note that ${\cal Y}$ is monotonously growing function of ${\cal E}$. The inequality $1+{\cal E}>0$ 
is always satisfied due to the
definition  (\ref{defcalE}). For {\em small\/} energy differences formula (\ref{Fcohmax}) goes to
\be
{\cal F}_{max}^{(coh)} \approx 1 -{\cal Y}^2/2 \approx 1 -{\cal E}^2/2.
\label{FEcohsmall}
\ee
Formulas (\ref{Fcohmax}) and (\ref{FEcohsmall}) were also derived in another way in \cite{DH}.

\section{Squeezed vacuum states}
\pst
Squeezed vacuum states are defined as follows,
\be
|\zeta\rangle =\exp\left(\frac12\left[\zeta \hat{a}^{\dagger 2} -\zeta^* \hat{a}^2\right]\right)
= \left(1-|\zeta|^2\right)^{1/4}\sum_{m=0}^{\infty}\frac{\sqrt{(2m)!}}{2^m m!}\zeta^m |2m\rangle,
\qquad |\zeta|<1.
\label{defsqz}
\ee
The fidelity between the states $|\zeta_1\rangle$ and $|\zeta_2\rangle$, and the mean energy in each state
are given by the formulas
\be
{\cal F} = \left[\frac{1 +\left|\zeta_1\zeta_2\right|^2 -|\zeta_1|^2 -|\zeta_2|^2}
{1 +\left|\zeta_1\zeta_2\right|^2 -2 \mbox{Re}\left(\zeta_1\zeta_2^*\right)}\right]^{1/2},
\qquad
E_j = \frac{1+|\zeta_j|^2}{2\left(1-|\zeta_j|^2\right)}.
\label{FEsqz}
\ee
The maximal fidelity is achieved for the complex numbers $\zeta_1$ and $\zeta_2$ having identical phases.
Therefore it is sufficient to consider real parameters $\zeta_1$ and $\zeta_2$ with $\zeta_1 \ge 0$.
Moreover, I assume that $\zeta_2>\zeta_1$, so that the relative energy difference ${\cal E}$ is postive
(otherwise one should simply interchange the states $|\psi_1\rangle$ and $|\psi_2\rangle$).
Resolving the equation
\be
{\cal E} =\frac{2\left(\zeta_2^2 -\zeta_1^2\right)}{\left(1+\zeta_1^2\right)\left(1-\zeta_2^2\right)}
\label{calEsqz}
\ee
with respect to $\zeta_2$ and putting the positive root 
\[
\zeta_2 =\sqrt{\frac{\zeta_1^2 +\tilde{\cal E}\left(1+\zeta_1^2\right)}{1 + \tilde{\cal E}\left(1+\zeta_1^2\right)}},
\qquad \tilde{\cal E}={\cal E}/2
\]
in the formula for the fidelity, one can arrive at the  expression 
\be
{\cal F}(z;\tilde{\cal E})= \frac{\sqrt{1 + \tilde{\cal E}\left(1+z\right)} +\sqrt{z^2 + z\tilde{\cal E}\left(1+z\right)}}
{(1+\tilde{\cal E})\left(1+z\right)}, \qquad z=\zeta_1^2.
\label{FzE}
\ee
For $z=1$ one obtains ${\cal F}(1;\tilde{\cal E})={\sqrt{1+2\tilde{\cal E}}}/(1+\tilde{\cal E})$, and one can see
after some algebra that the equation ${\cal F}(z;\tilde{\cal E})={\cal F}(1;\tilde{\cal E})$ can be transformed to
the equation $\tilde{\cal E}^2(1+z)^2(1-z)^2$, which has the unique admissible solution $z=1$.
On the other hand, ${\cal F}(0;\tilde{\cal E})=(1+ \tilde{\cal E})^{-1/2}< {\cal F}(1;\tilde{\cal E})$
for $\tilde{\cal E}>0$.
 Consequently,
${\cal F}(1;\tilde{\cal E})$ is the {\em maximal\/} value of fidelity for the given positive value $\tilde{\cal E}$. 
Rewriting this maximal value in terms of the symmetric relative energy difference (\ref{defY})
(in order to remove the restriction by positive energy differences), one can arrive at
the formula 
\be
{\cal F}_{max}^{(sqz)}= \left(1 +{\cal Y}^2/4\right)^{-1/2}.
\label{Fmaxsqz}
\ee
For ${\cal Y} \ll 1$ this formula goes to
\be
{\cal F}_{max}^{(sqz)} \approx 1 - {\cal Y}^2/8.
\label{Fmaxsqzapr}
\ee
It was shown in \cite{DH}) that formula (\ref{Fmaxsqz}) gives in fact the maximal fidelity for {\em arbitrary\/}
squeezed coherent (i.e., pure Gaussian) states (when arbitrary displacements can be added).

\section{Negative binomial states and their special cases}
\pst
Now let us consider the family of ``negative binomial states''
 \begin{equation}
|\zeta,\mu,\{\vartheta_n\}\rangle=\sum_{n=0}^{\infty} e^{i\vartheta_n}
\left[(1-\zeta)^{\mu}\frac{\Gamma(\mu+n)}{\Gamma(\mu)n!}\zeta^n\right]^{1/2}
\:|n\rangle, \qquad 0\le\zeta<1, \quad \mu >0. 
\label{negbin}
\end{equation}
These states were apparently introduced under different names (although not in the most general case) 
in \cite{AharLer71,OnoPa,DMM74}. The name was given in
\cite{JoLa89,Matsuo90}. After that, the states (\ref{negbin}) attracted some attention for the past two decades
\cite{Wang00,Liao01,OMD03,Abdal07}, 
in particular, because they go to the coherent states in the limit $\mu\to\infty$, $\zeta\to 0$
(provided the product $\mu\zeta$ is maintained fixed) and to the so-called coherent phase
states \cite{Ler70,ShapShep91,Sudar93,DM95,Vourd96,Wun01}  for $\mu=1$ (if $\vartheta_n=n\vartheta_0$). 
The mean energy of the state
(\ref{negbin}) does not depend on the phases $\vartheta_n$:
\be
\langle E_{\zeta}\rangle = \frac12 + \frac{\mu\zeta}{1-\zeta}.
\label{meanE}
\ee
But the fidelity between different states does depend on the differences in phases. Obviously, the fidelity
is maximal for the fixed values $\zeta$ and $\zeta'$ of two states if all these phase differences equal zero.
For this reason we consider here only the special case of $\vartheta_n \equiv 0$, thus omitting the parameters
$\{\vartheta_n\}$.
Then the fidelity and the energy difference $\Delta E=\langle E_{\zeta'}\rangle - \langle E_{\zeta}\rangle$ 
between the states $|\zeta,\mu\rangle$ and $|\zeta',\mu\rangle$ read
\be
{\cal F} = \frac{\left[(1-\zeta)(1-\zeta')\right]^{\mu}}
{\left(1-\sqrt{\zeta\zeta'}\right)^{2\mu}},
\qquad
\Delta E = \frac{\mu(\zeta' -\zeta)}{(1-\zeta)(1-\zeta')}.
\label{FE-negbin}
\ee
For the fixed difference $\zeta' -\zeta$ the energy difference can be arbitrarily big if
$\mu\to\infty$ or $\zeta\to 1$. Therefore we consider again the relative energy difference
\be
{\cal E} \equiv \Delta E /\langle E_{\zeta}\rangle = 
\frac{2\mu(\zeta' -\zeta)}{(1-\zeta')\left[1+(2\mu-1)\zeta\right]},
\label{calE}
\ee
which is certainly limited.
Resolving equation (\ref{calE}) with respect to $\zeta'$, one obtains the following expression for the
fidelity as function of ${\cal E}$, $\zeta$ and $\mu$:
\be
{\cal F}({\cal E}; \zeta; \mu) = \left[\frac{\sqrt{1 + {\cal E}_{\mu}g(\zeta,\mu)} +
\sqrt{\zeta^2 + \zeta{\cal E}_{\mu}g(\zeta,\mu)}
}
{1 + \zeta+ {\cal E}_{\mu}g(\zeta,\mu)}
\right]^{2\mu},
\label{FxiE}
\ee
where
\[
{\cal E}_{\mu}= {\cal E}/(2\mu), \qquad
g(\zeta,\mu) = 1 + \varkappa\zeta, \qquad \varkappa =2\mu-1.
\]
Attempts to find the maximal value of function (\ref{FxiE}) by calculating the derivative $\partial {\cal F}/\partial\zeta$
lead to complicated algebraic equations, which are difficult to analyze. Fortunately, the parameters of function
(\ref{FxiE}) are adjusted in such a way that a simpler approach is possible.

Let us look for the values of $\zeta$ that give some chosen value $\beta$ of the fraction inside the square brackets in (\ref{FxiE}).
This is equivalent to the equation
\[
\sqrt{\zeta^2 + \zeta{\cal E}_{\mu}g} = \beta\left(1 + \zeta+ {\cal E}_{\mu}g\right) -
\sqrt{1 + {\cal E}_{\mu}g}.
\]
Taking the squares of each side of this equation, one can see that it can be reduced to the form
\[
2\beta\left(1 + \zeta+ {\cal E}_{\mu}g\right) \sqrt{1 + {\cal E}_{\mu}g}=
\left(1 + \zeta+ {\cal E}_{\mu}g\right)\left[\beta^2\left(1 + \zeta+ {\cal E}_{\mu}g\right) +1-\zeta \right].
\]
Consequently, the common term $\left(1 + \zeta+ {\cal E}_{\mu}g\right)$ in both sides can be canceled, so that after
taking again the squares of both sides one arrives at the {\em quadratic\/} equation. Moreover, this quadratic equation
has rather specific form:
\be
A^2 \zeta^2 +2B\zeta + C^2=0,
\label{eq}
\ee
\[
A= \beta^2\left(1+{\cal E}_{\mu}\varkappa\right) -1, \qquad
C= \beta^2\left(1+{\cal E}_{\mu}\right) -1, \qquad
B= \beta^4\left(1+{\cal E}_{\mu}\varkappa\right)\left(1+{\cal E}_{\mu}\right)
-\beta^2 {\cal E}_{\mu}(1+\varkappa) -1.
\]
The discriminant of this equation goes to zero if $B=\pm AC$. In this case equation (\ref{eq}) has a {\em single\/}
solution, which corresponds to the only extremum of function (\ref{FxiE}).
Taking $B=AC$, one arrives at the relation ${\cal E}_{\mu}^2(\zeta\varkappa +1)^2=0$, which can be satisfied for 
$\mu>0$ and $0\le\zeta<1$ only for ${\cal E}=0$. This corresponds to the maximal fidelity ${\cal F}=1$, so this
solution is trivial. Nontrivial results arise for $B=-AC$:
\[
\beta_*^2 = \frac{1+{\cal E}_{\mu}(1+\varkappa)}{\left(1+{\cal E}_{\mu}\right)\left(1+{\cal E}_{\mu}\varkappa\right)},
\qquad
A= \frac{{\cal E}_{\mu}\varkappa}{1+{\cal E}_{\mu}}, 
\qquad C= \frac{{\cal E}_{\mu}}{1+{\cal E}_{\mu}\varkappa}, 
\qquad B = - AC.
\]
The unique solution of (\ref{eq}) is
\be
\zeta_*= -\frac{B}{A^2}= \frac{C}{A} = \frac{1+{\cal E}_{\mu}}{\varkappa\left(1+{\cal E}_{\mu}\varkappa\right)}.
\label{solzeta}
\ee

\subsection{States with $\mu>1$}
The further results depend on the value of parameter $\mu$ (or $\varkappa$). Let us consider first the case $\varkappa>0$,
assuming that ${\cal E}>0$ (otherwise  parameters $\zeta$ and $\zeta'$ should be interchanged). Then one can see
that the condition $\zeta_*<1$ is equivalent to $(\varkappa -1)(1+{\cal E})>0$, i.e., $\varkappa >1$ or $\mu>1$
(since $1+{\cal E}>0$ always). In this case function (\ref{FxiE}) has the only extremum 
\be
\beta_*^{2\mu}=\left[\frac{1+{\cal E}}{\left(1+{\cal E}_{\mu}\right)\left(1+{\cal E}_{\mu}\varkappa\right)}\right]^{\mu}
= \left[\frac{1+{\cal E}}{\left(1+{\cal E} +{\cal E}^2(2\mu -1)/(2\mu)^2\right)}\right]^{\mu}
\label{beta*2mu}
\ee
inside the interval $0\le \zeta<1$. This extremum is a {\em maximum\/}, as can be easily seen by comparing it with
the values of ${\cal F}(\zeta)$ at the points $\zeta=0$ and $\zeta=1$:
\be
{\cal F}(\zeta=0) = \left(1+{\cal E}_{\mu}\right)^{-\mu},
\qquad
{\cal F}(\zeta=1) = \left(\frac{\sqrt{1+{\cal E}}}{1+{\cal E}/2}\right)^{2\mu}.
\label{F01}
 \ee
Replacing  ${\cal E}$ in (\ref{beta*2mu}) by its expression in terms of the symmetrical relative energy difference ${\cal Y}$,
\be
{\cal E} = {\cal Y}\sqrt{1 + {\cal Y}^2/4} + {\cal Y}^2/2, \qquad
{\cal E}^2 = {\cal Y}^2(1+{\cal E}),
\label{E-Y}
\ee
one can arrive at the formula
\be
{\cal F}_{max}^{(negbin)}= \left(1+\frac{2\mu -1}{4\mu^2}{\cal Y}^2\right)^{-\mu},
\label{FnegY}
\ee
which is valid for $\mu\ge 1$, both for positive and negative values of ${\cal E}$.
In the limit $\mu\to\infty$ formula (\ref{FnegY}) goes to (\ref{Fcohmax}),
as one can expect.
The Taylor expansion of (\ref{FnegY}) reads
\be
{\cal F}_{max}^{(negbin)} \approx 1 - \frac{2\mu-1}{4\mu}{\cal Y}^2, \quad
\mbox{if}\;\; {\cal Y} \ll 1 \;\;\mbox{and}\;\; \mu\ge 1.
\label{Fmaxbigmu}
\ee
In the important special case of $\mu =1$
(``coherent phase states'' whose photon statistics coincides with Planck's distribution),
the general formulas assume the following forms:
\be
{\cal F}_{max}^{(phase)}= \left(1 +{\cal Y}^2/4\right)^{-1}, \qquad
{\cal F}_{max}^{(phase)} \approx 1 - {\cal Y}^2/4 \quad \mbox{for}\; {\cal Y} \ll 1.
\label{Fmaxfaz}
\ee

\subsection{The case of $\mu<1$. Hyper-Poissonian states.}

If $\mu<1$, the extremal point $\zeta_*$ given by (\ref{solzeta}) does not belong to the allowed interval
$0\le\zeta <1$. Consequently, the function (\ref{FxiE}) varies monotonously with $\zeta$, 
and comparing two values in (\ref{F01}) one can verify that the maximal value is attained for $\zeta=1$.
Therefore
\be
{\cal F}_{max}^{(negbin)}= \left(1+{\cal Y}^2/4\right)^{-\mu}, \qquad \mu \le 1,
\label{FnegYzeta1}
\ee
\be
{\cal F}_{max}^{(negbin)} \approx 1 - \mu{\cal Y}^2/4, \qquad {\cal Y} \ll 1, \qquad \mu <1.
\label{Fmaxsmalmu}
\ee
If $\mu=1/2$, then formula (\ref{FnegYzeta1}) coincides with formula (\ref{Fmaxsqz}) for vacuum squeezed states
(although the statistics of these states are different, and fidelities for arbitrary parameters, different from 
$\zeta=1$, do not coincide). This specific value of $\mu$ has an interesting physical meaning, namely, it is
a boundary between ``usual super-Poissonian'' states and ``hyper-Poissonian'' ones \cite{apsi}.
To understand the difference, 
remember that the statistics of squeezed vacuum states is called
sometimes ``super-chaotic'' \cite{Walls75,Sot81},
since Mandel's $Q$-factor in these states equals $1+2\langle n\rangle$, i.e., it is twice bigger
(for $\langle n\rangle \gg 1$) than its value $Q_{therm}=\langle n\rangle$
in ``usual chaotic'' thermal states.
On the other hand, it was shown in \cite{apsi} that the mean number
of quanta $\langle n_{-}\rangle$ in the ``photon-subtracted'' state $|\psi_-\rangle = \hat{a}|\psi\rangle$
is {\em bigger\/} than the mean number
of quanta $\langle n_{+}\rangle$ in the ``photon-added'' state $|\psi_+\rangle = \hat{a}^{\dagger}|\psi\rangle$, 
if Mandel's parameter of the initial state
$|\psi\rangle$ obeys the inequality $Q>1+2\langle n\rangle$.
Negative binomial states belong to this class if
 $\mu<1/2<\xi(1-\mu)$, since
$Q={\xi}/{(1-\xi)}$ and $\langle{n}\rangle=\mu{Q}$.

 If $\mu \ll 1$, then  the maximal relative energy difference between two negative binomial states 
 can be very big, despite that they could seem very ``close'' from the point of view of the fidelity,  since
the inversion of formula (\ref{Fmaxsmalmu}) for the fixed fidelity reads ${\cal E}_{max} = 2\sqrt{(1-{\cal F})/\mu}$.
Such an effect is explained by the properties of superpositions (\ref{negbin}) with $\mu\ll 1$ and $1-\zeta\ll 1$:
although the probability of the vacuum state is very close to unity in these states, the number (photon) distribution 
function has a very long ``tail''
of states with very small but very slowly decaying probabilities $p_n \approx \mu^2/n$.

\section{Binomial states}
\pst
The last example is related to the {\em binomial states}, which are intermediate between the Fock
and coherent states \cite{Ahar73,Stol85,Lee85,Vid94}:
\be
|p,M\rangle = \sum_{n=0}^M \left[\frac{M!}{n!(M-n)!}p^n(1-p)^{M-n}\right]^{1/2}|n\rangle.
\label{defbin}
\ee
Here $M$ is positive integer and $0\le p \le 1$. We take all phases of the amplitude coefficients
equal to zero, for the same reasons as it was done in the preceding sections. The fidelity and relative
energy difference between two states with parameters $p$ and $p'$ are as follows,
\be
{\cal F} =\left[\sqrt{pp'} +\sqrt{(1-p)(1-p')}\right]^M,
\qquad
{\cal E} = \frac{M(p'-p)}{1/2 +Mp}.
\label{FEbin}
\ee
The same approach as used in the preceding section
\footnote{Namely, one should put $p'(p,{\cal E},M) = p(1+{\cal E}) +{\cal E}/(2M)$ in the first formula of (\ref{FEbin}) 
and look for the condition of
existence of the unique solution of the equation $\sqrt{pp'} +\sqrt{(1-p)(1-p')}=\beta$. This problem can be
reduced again (after some algebra) to solving {\em quadratic\/} equations --- of course, this is a great ``present of nature''
due to the specific structure of the binomial states. The maximum value is attained for 
$p_*=(2M-{\cal E})/[4M(M+1)(1+{\cal E})]$, and one can verify that $0\le p_* \le 1$ for any value of $M$.}
 leads to the following result:
\be
{\cal F}_{max}^{(bin)} =\left(1 -\frac{2M+1}{4M^2}{\cal Y}^2\right)^M.
\label{Fmaxbin}
\ee
The expression inside the brackets cannot be negative, because the maximal relative energy change is limited:
according to (\ref{FEbin}), $-2M/(1+2M)\le{\cal E}\le 2M$,  so that ${\cal Y} \le 2M/\sqrt{2M+1}$ for any sign of ${\cal E}$.
Note that formally formula (\ref{Fmaxbin}) can be obtained from (\ref{FnegY}) simply by 
means of the substitution $\mu=-M$.
For small energy difference,
\be
{\cal F}_{max}^{(bin)} \approx 1 -\frac{2M+1}{4M}{\cal Y}^2, \qquad {\cal Y} \ll 1.
\ee
If $M\to \infty$, formula (\ref{Fmaxbin}) goes to the formula for the coherent states (\ref{Fcohmax}).

\section{Summary}
\pst
We have succeeded in obtaining exact formulas for the maximal fidelity of two quantum states belonging 
to various popular families, for the fixed relative energy difference between the states. Of course,
this happened due to an amazing ``fine tuning'' of coefficients in the corresponding finite and infinite
superpositions of the Fock states. In more general cases the upper bounds can be found only numerically.
A practical consequence of the performed study is the indication that only rather high levels of fidelity
can guarantee a ``real'' closeness of two states {\em in the absence of any additional information}.
For example, if it is known that two states are coherent, and one of them has the parameter $\alpha_1=1/\sqrt{2}$
(when $E_1=1$), the fidelity $90$\% means, according to (\ref{cohFE}), that  
the minimal and maximal values of $\alpha_2$ can be  $0.38$ and $1.03$. The corresponding relative energy differences
are equal to $35$\% and $56$\%, respectively. If one wishes to reduce the maximal possible relative energy difference to
the level of $10$\%, the fidelity between two {\em unknown\/} coherent states must be higher than $0.995$.
For squeezed and especially negative binomial states with $\mu< 1/2$, the requirements can be much stronger.

\section*{Acknowledgment}
\pst
The author acknowledges the partial financial support provided by
the Brazilian agency CNPq.

\end{document}